\documentclass[a4paper,prl,twocolumn,showpacs,amssymb,apscolor,superscriptaddress,psfig,preprintnumbers]{revtex4}
\usepackage{graphicx}
\usepackage{color,amsmath,graphicx,latexsym,psfig}
\usepackage{epsfig}
\usepackage{dcolumn}
\usepackage{bm}
\usepackage{xspace}
\usepackage{placeins,soul}

\newcommand {\refa}  [1] {\mbox{Fig.\,\ref{#1}}}

\newcommand{\etal}{{\it et al.}}

\newcommand{\musr}[0]{$\mu$SR}

\newcommand{\li}{Li$_{y}$VO$_{x}$-NT}

\newcommand{\livox}{Li$_{0.1}$VO$_{x}$-NT}
\newcommand{\vox}{VO$_{x}$-NTs}

\begin{document}

\title{High temperature ferromagnetism of Li-doped vanadium oxide nanotubes
\footnote{with some amendments published in

Europhysics Letters (EPL) {\bf 88} (2009) 57002

http://epljournal.edpsciences.org/}}


\author{A.I.\ Popa}
\affiliation{Institute for Solid State Research, IFW Dresden, 01171 Dresden, Germany}
\author{E. Vavilova}
\affiliation{Institute for Solid State Research, IFW Dresden, 01171 Dresden, Germany} \affiliation{Zavoisky Physical-Technical Institute of
RAS, 420029 Kazan, Russia}
\author{Y.C.\ Arango}
\affiliation{Institute for Solid State Research, IFW Dresden, 01171 Dresden, Germany}
\author{V. Kataev}
\email[]{v.kataev@ifw-dresden.de} \affiliation{Institute for Solid State Research, IFW Dresden, 01171 Dresden, Germany}
\author{C. T\"aschner}
\affiliation{Institute for Solid State Research, IFW Dresden, 01171 Dresden, Germany}
\author{H-H. Klauss}
\affiliation{IFP, TU Dresden, D-01069 Dresden, Germany}
\author{H. Maeter}
\affiliation{IFP, TU Dresden, D-01069 Dresden, Germany}
\author{H. Luetkens}
\affiliation{Laboratory for Muon-Spin Spectroscopy, Paul Scherrer Institut, CH-5232 Villigen PSI, Switzerland}
\author{B.\ B\"uchner} \affiliation{Institute for Solid State Research, IFW Dresden, 01171 Dresden, Germany}
\author{R.\ Klingeler}
\affiliation{Institute for Solid State Research, IFW Dresden, 01171 Dresden, Germany}

 \pacs{75.20.-g; 75.75.+a; 73.22.-f; 76.60.-k; 76.75.+; 76.30.-v}

\begin{abstract}
The nature of a puzzling high temperature ferromagnetism of doped mixed-valent vanadium oxide nanotubes reported
earlier by Krusin-Elbaum {\it et al.}, {\it Nature} {\bf 431} (2004) 672, has been addressed by static magnetization, muon spin relaxation, nuclear
magnetic and electron spin resonance spectroscopy techniques. A precise control of the charge doping was achieved by electrochemical Li
intercalation. We find that it provides excess electrons, thereby increasing the number of interacting magnetic vanadium sites, and, at a certain
doping level, yields a ferromagnetic-like response persisting up to room temperature. Thus we confirm the surprising previous results on
the samples prepared by a completely different intercalation method. Moreover our spectroscopic data provide first ample evidence for the {\it bulk}
nature of the effect. In particular, they enable a conclusion that the Li nucleates superparamagnetic nanosize spin clusters around the
intercalation site which are responsible for the unusual high temperature ferromagnetism of vanadium oxide nanotubes.
\end{abstract}

\maketitle

The quest for new materials with novel physical properties and functionalities continuously pushes forward materials's science efforts to structure
'old known' bulk solids into nanoscaled low-dimensional objects (see, e.g.,~\cite{nanostructures}). The magnetic and electronic properties of such
nanostructures are mainly governed by their surfaces and interfaces which renders them promising materials for tailored functionalization. For oxide
heterostructures the great potential of such an approach is, e.g., highlighted by recent findings of novel effects at interfaces between nonmagnetic
oxide insulators where magnetism, metallic behavior and even superconductivity have been reported~\cite{interfaces}. In order to tailor
nanostructured functional elements, different design strategies are being used, such as fabrication of multilayers, heterostructures with quantum
confinement (quantum wells, wires and dots) etc. Another promising approach is based on self-assembly of elementary nano- and even atomic-size
building blocks. In particular, two-dimensional layers of transition metal (TM) oxides rolled up into nanosized multiwalled tubes or scrolls attract
a rapidly growing attention owing to unique physical properties not occurring in the bulk parent materials (see, e.g.,~\cite{Wang06}). A recent
example of such a nanostructured TM oxide is multiwalled vanadium oxide nanotubes VO$_x$-NTs~\cite{Krumeich99,Worle02}. Vanadium ions are mixed
valent in this material with an average valence count of $\sim +4.4$ (i.e. $x\approx 2.2$)~\cite{Liu,Hellmann08}. This results in a roughly equal
number of magnetic V$^{4+}$ ($3d^1, S = 1/2$) and nonmagnetic V$^{5+}$ ($3d^0, S = 0$) sites in the VO$_x$ layers. The spins associated with the
former sites act either as individual spins or strongly gapped antiferromagnetic (AF) dimers or trimers~\cite{Vavilova}. As reported by
Krusin-Elbaum {\it et al.} ~\cite{Krusin}, doping of VO$_x$-NTs with either holes or electrons via iodine or lithium intercalation yields a nonlinear and hysteretic magnetization response to applied magnetic fields suggesting the occurrence of ferromagnetism that persists even
above room temperature. Such high temperature ferromagnetism (HTFM) is very surprising and unexpected for an oxide material comprising TM
ions with a small spin-1/2 and provides a remarkable example of novel functionalities in nanostructured oxides.

Since a conclusion on the HTFM in the doped \vox\  is based in Ref.~\cite{Krusin} only on the results of the SQUID magnetometry, where
artefacts due to uncontrollable ferromagnetic impurities cannot be always excluded, we have conducted a new experimental study using a combination of
different techniques. Our objectives were: (i) to obtain an independent evidence of HTFM in Li doped \vox, and, if confirmed, (ii) to obtain insights
into the origin of this unusual phenomenon. To achieve these goals we have prepared a series of Li-\vox\ samples by a completely different
electrochemical method of Li intercalation that as compared with the chemical method used in Ref.~\cite{Krusin} enables an accurate control of the
doping level. We have performed a complex experimental investigation of these samples with three different local spin probe techniques,
namely, electron spin resonance (ESR), nuclear magnetic resonance (NMR) and muon spin relaxation/rotation (\musr) spectroscopies, along with
measurements of the bulk static magnetization. We find that for a particular concentration of the Li-dopant a large magnetization $M$ which can be
easily saturated even at room temperature by a magnetic field $\mu_0 H$ of about 1\,T is present in the sample. Thereby we reproduce and
confirm the results of Ref.~\cite{Krusin} on the samples prepared by a different method. Moreover, \musr\ and NMR measurements give evidence for a
bulk nature of the effect. NMR data suggest that the magnetization is not uniform throughout the sample and that strongly magnetic regions are formed
around the intercalated Li sites. ESR experiments reveal a sharp signal that bears essential features of a superparamagnetic resonance. We argue that
Li intercalation affects the charge disproportionation and hence the spin states and magnetic interactions in the rolled-up VO$_x$ layers thereby
promoting, for particular doping levels, the formation of nanosized interacting spin clusters that behave similar to superparamagnetic nanoparticles.

Pristine multiwalled vanadium oxide nanotubes were synthesized by
a hydrothermal technique, previously described in~\cite{Liu}.
Electrochemical treatment that enables a precise control of the Li
doping was done by means of two-electrode Swagelok-type cells,
each including 83.3 wt\% of active material (VO$_{x}$-NT), 16.7
wt\% of Carbon SP (Timcal), a Li metal anode, and 1:1 ethylene
carbonate/dimethyl carbonate solution of 1M LiPF$_{6}$
(LP30-Ferro) that was used as electrolyte. The electrochemical
doping was done using a VMP controller (Ametek Princeton Applied
Research), in galvanostatic mode, with a discharge rate of 1 Li
per formula unit in 100\,h. By applying the constant current mode
Li ions have been intercalated to the active material until a
desired composition Li$_{y}$VO$_{x}$-NT was achieved: For the
present study samples with $y$ = 0, 0.05, 0.10, 0.15 and 0.6 have
been prepared. For all samples, magnetization has been measured in
the temperature range $T = 2 - 350$\,K and in fields up to 5 T by
means of a Quantum Design MPMS. $^{51}$V NMR data were recorded on
a Tecmag pulse solid-state NMR spectrometer at $T = 4.2 - 300$\,K.
The NMR spectra were acquired by a point-by-point magnetic field
sweeping. For ESR experiments we used an X-band EMX Bruker ESR
spectrometer operational at $T = 3.5 - 300$\,K. Muon spin
relaxation ($\mu$$^{+}$SR) data on undoped and doped VO$_{x}$-NTs
were obtained at the GPS spectrometer of PSI.

\begin{figure}[t]
 \centering
 \includegraphics[width=0.8\columnwidth,clip]{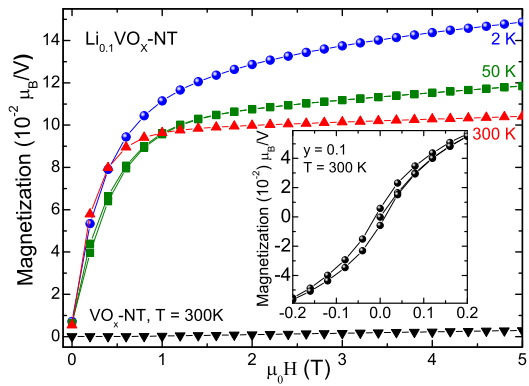}
 \caption{$H$-dependence of the magnetization of Li$_{y}$VO$_{x}$-NT with $y=0.1$ ($T$ =
 300\,K, 50\,K, 2\,K) and $y=0$ ($T$ = 300\,K). Inset: Part of the $M(H)$
 loop, at 300\,K, for $y=0.1$.}
 \label{MH}
\end{figure}

Our static magnetic data show an increase of the magnetization in
Li$_{y}$VO$_{x}$-NT with $0 < y \leq 0.6$ as compared with the
undoped material. This observation is consistent with a higher
number of paramagnetic V sites suggested by the NMR data (see
below). Note, that a spin-gap feature in the $T$-dependence of the
susceptibility in pristine VO$_x$-NT attributed to the presence of
dimers~\cite{Krusin,Vavilova} vanishes upon Li doping. The main
result of the magnetization study is the observation of a
non-linear behavior for the doping level $y=0.1$ which strongly
differs from the magnetic response at other Li contents. At
300\,K, the $M(H)$ curve practically saturates in a field of $\sim
1$\,T at a value of $\sim 0.1\mu_{\rm B}$ per V site with a small
subsequent linear increase at higher fields (see ~\refa{MH}).
The data at 50\,K and 2\,K (\refa{MH}) illustrate that at low
$T$ this saturation is superimposed by a stronger linear
contribution to $M(H)$. Note, that the data exhibit a small
hysteresis with the coercivity $2H_c\sim 23$\,mT  at 300\,K (see ~\refa{MH}, inset).

\begin{figure}[t]
 \centering
 \includegraphics[width=0.75\columnwidth]{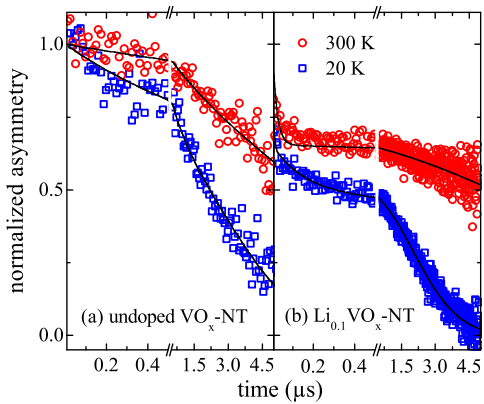}
 \caption{{Selected ZF-\musr\ spectra at 300\,K and 20\,K of \li\ with (a) $y=0$ and (b) $y=0.1$.}
 The solid lines denote the evaluation of the data with a function as described in the text.
 Note the change of the timescale at 0.5\,$\mu$ s.}
 \label{musr}
\end{figure}

Further insight into the magnetic properties is provided by zero
field (ZF) \musr\ data. Selected ZF-\musr\ spectra obtained on
\li\ with $y=0$ and $y=0.1$, i.e. for the pristine and the
magnetic materials, (see ~\refa{musr}). For $y=0$, both at
300\,K and at 20\,K the data show only a small decrease of the
asymmetry signal $A(t)$ at short times. This clearly implies the
absence of magnetic order in this temperature range. On a longer
time scale of $\mu$s, there is a decrease due to the slow
relaxation which we attribute to nuclear and fast fluctuating
electronic magnetic moments.

The most important result of the ZF-\musr\ study on \livox\ is the
observation of a significant and rapid loss of asymmetry at early
times. As displayed (see ~\refa{musr}(b)), most of the
relaxation occurs already during the dead time of the spectrometer
($\approx$ 5 ns). The full asymmetry scale was defined by a
subsequent measurement of a nonmagnetic compound. Such a rapid
relaxation clearly indicates that a significant fraction of the
muons experiences a local quasi-static magnetic field. The absence
of an oscillating signal proves a broad static magnetic field
distribution within the compound.

Assuming for a quantitative analysis that a fraction of ${a}$ muons
experiences both the nuclear and the a electronic static magnetic field,
the data are described by $A(t)=(1-a)\cdot{}e^{-0.5(\sigma_{nuc}t)^2} +
a\cdot{}(2/3e^{-\lambda_Tt}+1/3e^{-\lambda_Lt})$. Here $\sigma_{nuc}$ and
$\lambda_{T,L}$ are muon relaxation rates due to interaction with nuclear
and electron spins, respectively ~\cite{Blundell}. This analysis implies
that $\sim$2/3 of the muons in \livox\ experience a static magnetic field
originating from a magnetic ordered volume fraction at or in the vicinity
of the muon stopping site(s)~\cite{fraction}. The \musr\ data hence
unambiguously proof the bulk, and {\em not} impurity related, character of
the magnetism found in the static magnetization.

\begin{figure}[t]
  \centering
  \includegraphics[width=0.9\columnwidth]{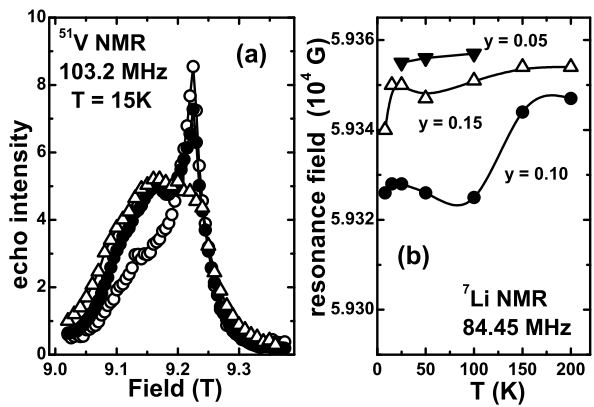}
  \caption{Low-$T$ $^{51}$V NMR spectra (a) and $T$-dependencies of
the resonance field of $^{7}$Li NMR (b) for \li : $y$\,=\,0
($\circ$), 0.05 ($\blacktriangledown$), 0.1 ($\bullet$) and 0.15
($\vartriangle$)). Lines are guides to the eye.}
  \label{nmr}
\end{figure}

NMR studies on $^{51}$V and $^{7}$Li shed further light on the
effect of Li doping on the local magnetic properties. The $^{51}$V
NMR data (see ~\refa{nmr}(a)) reveal a gradual increase of the
relative intensity of the low-field fast-relaxing shoulder of the
$^{51}$V signal at $\sim 9.16$\,T upon doping. Since this part of
the spectrum is associated with the response of the magnetic ions'
nuclei, shifted due to hyperfine interaction from the central
slow-relaxing nonmagnetic $V^{5+}$ peak at $\sim 9.22$\,T
~\cite{Vavilova}, its growth indicates an increasing fraction of
magnetic vanadium ions upon Li intercalation and confirms that the
doping process affects the whole sample. Though magnetic ordering
usually creates a shift or splitting of the NMR spectrum, neither
was observed for the $^{51}$V signal for the \livox\ sample in a
sufficiently wide field range. The absence of the significant loss
of its intensity suggests that the large part of V ions does sense
the charge doping but not the internal field.

The situation with $^{7}$Li NMR is different: The resonance field
of the signal steadily changes with doping. However, the line
shift for $y=0.1$ is much larger compared to $y=0.05$ and $y=0.15$
(see ~\refa{nmr}(b)). This suggests the presence of an internal
field at the Li sites only for the strongly magnetic \livox . The
absence of a second unshifted $^{7}$Li line at the expected place
between the line positions of Li$_{0.05}$ and Li$_{0.15}$ samples
indicates that practically {\it all} Li nuclei experience internal
magnetic fields.

The ESR study of undoped VO$_{x}$-NTs at a frequency $\nu =
9.5$\,GHz reveals, similar to  other results~\cite{Kweon}, a
spectrum comprising two overlapping resonance lines with slightly
different $g$-factors of $\sim 2.0$ and $\sim 1.97$ (not shown).
The first line can be assigned to quasi-free spins associated with
V$^{4+} (S = 1/2)$ ions in the tetrahedral position. The second
line is due to V$^{4+}$ ions in the distorted octahedral
coordination which are coupled magnetically into dimers and
trimers~\cite{Vavilova}.

\begin{figure}[t]
  \centering
  \includegraphics[width=1.0\columnwidth{}]{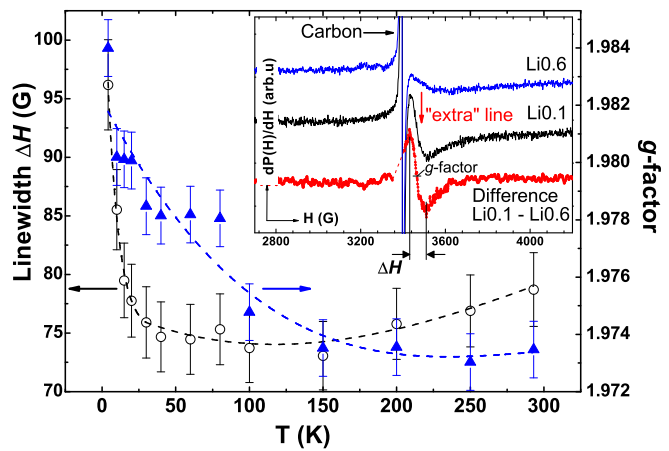}
  \caption{Inset: ESR spectra (field derivative of absorption)
   of \li\ for $y = 0.6$ (top) and 0.1 (middle) at $T = 300$\,K.
  The "extra" narrow line which is
specific for the strongly magnetic sample with $y = 0.1$ (marked
by the arrow in the middle spectrum) is singled out by subtracting
the top from the middle spectrum (bottom curve). The central sharp
line is due to the carbon additive. Main panel: $T$-dependence of
the width $\Delta H$ and the $g$-factor of the "extra" line in the
ESR spectrum of $y = 0.1$. Dash lines are guides for the eye.}
\label{esr}
\end{figure}

In the samples of Li$_{y}$VO$_{x}$-NTs with $y = 0.05, 0.1, 0.15$
and 0.6 a systematic evolution of the lines in the ESR spectrum
with increasing the Li doping gives evidence for the increasing
number of magnetic V$^{4+}$ ions, and a reduction of the
contribution related to spin dimers and trimers, in agreement with
the magnetization data (see above). We note that the carbon black
used as a conductive additive for the preparation of Li-doped
samples partially screens the penetration of microwaves into the
interior of the sample thereby reducing the signal-to-noise ratio
and superimposes an additional ESR line (see ~\refa{esr},
inset). The latter can be identified by measuring the carbon
additive separately and accurately subtracted from the spectra.
The central result of the ESR study is the observation of an
'extra' narrow resonance line in the strongly magnetic  $y = 0.1$
sample. Despite the disturbing effect of the carbon line, the
'extra' line is clearly visible in the raw data (see
~\refa{esr}, inset, middle curve) and can be accurately singled
out, e.g., by subtracting from the $y = 0.1$ spectrum the spectrum
of a weakly magnetic sample with a different Li content (see
~\refa{esr}, inset, bottom curve). The width of this 'extra'
line $\Delta H \sim 70$\,G is significantly smaller compared to
the signals from paramagnetic V$^{4+}$ ions whereas the $g$-factor
of $\sim 1.98$ is similar. Both $\Delta H$ and $g$ change little
in a broad $T$-region above $\sim 40 - 70$\,K and become
temperature dependent at lower $T$ (see ~\refa{esr}). As will
be discussed below, this narrow signal is directly related with
the ferromagnetic-like behavior of the $y = 0.1$ sample.

Uniform long-range magnetic order in the low-spin ($S = 1/2$)
vanadium oxides is not competitive at high temperatures with
thermal fluctuations and/or insulator-to-metal transitions (see,
e.g.,~\cite{Goodenough}). A ferromagnetic-like magnetization up to
room and even higher temperatures is in particular unexpected.
However, in the case of \vox\ an alternative cluster scenario of
ferromagnetism is clearly corroborated by our  \musr\ and NMR
results. \musr\ data yield about 2/3 of the magnetic volume
fraction and a broad distribution of static magnetic fields within
the sample. The absence of an appreciable shift of the $^{51}$V
NMR signal in the ferromagnetic sample suggests that it originates
from the regions outside ferromagnetic clusters. The $^{51}$V-NMR
line shift from the nuclei inside the clusters is expected to be
as big as in magnetically ordered compounds, hence displacing the
signal out of the observation range (see e.g.,~\cite{Kiyama}).
Furthermore, owing to a possible distribution of the clusters'
size, as also suggested by the $\mu$SR data, this signal could be
very broad and thus practically undetectable. On the other hand,
the $^7$Li-NMR data give clear evidence that ferromagnetism is
confined to regions around the lithium site which yields a shift
of the $^7$Li-NMR line by the internal magnetic field inside the
cluster.

Thus one can argue that the Li which is intercalated between the vanadium oxide layers in the walls of \vox\ ~\cite{Hellmann08} may play a role of
nucleating centers for spin clusters of different size. In fact, the magnetization data (\refa{MH}) bear features characteristic of a magnetic
response of superparamagnetic particles with a broad size distribution, such as, e.g., samples of ultra-fine $\gamma$-Fe$_2$O$_3$ particles studied
in ~\cite{Coey72}. Specifically, the small hysteresis in $M(H)$ at room temperature suggests the presence of large clusters with a blocking
temperature $T_{\rm B} > 300$\,K. At higher external fields they are already saturated and the observed finite slope of $M(H)$ in this field regime
is mainly determined by smaller unblocked superparamagnetic clusters ($T_{\rm B} \ll 300\,K$)~\cite{Coey72}. The sharp ESR line with the spin-only
paramagnetic resonance field $H_{res}^{par} = h\nu/g\mu_B$ observed in the strongly magnetic \livox\ sample (~refa{esr}) can be straightforwardly
assigned to a resonance response of those small unblocked clusters. Above $T_{\rm B}$ the anisotropy field, that otherwise produces a shift and
broadens the signal, is averaged due to thermal fluctuations yielding a narrow line at $H_{res}^{par}$ (see, e.g.,~\cite{Berger}). At low
temperatures one can consider the shift of the effective $g$-factor from the spin-only value and the increase of the width of this signal (see ~\refa{esr}) as an indication of approaching the $T_{\rm B}$ of the resonating superparamagnetic clusters. According to the study of the effect of
the particle size on the ESR response in ~\cite{Gazeau}, for small particle size ($\approx 5$\,nm) the spectrum is defined by an isotropic and
unshifted narrow line. In contrast, for a large particle size ($\approx 10$\,nm) the associated anisotropy field is much stronger, therefore, thermal
fluctuations even on a room temperature scale cannot overcome the anisotropic orientation of the magnetic moments. As a result, the spectrum is broad
and shifted towards lower fields due to the influence of the remaining orientational anisotropy. Thus, in the case of a broad size distribution, the
ESR response of the blocked clusters in \livox\ could be smeared out and become unobservable, in particular, also due to the limited sensitivity
caused by the carbon additive (see above).

Obviously, local charge and structural distortions around an
intercalated Li$^{+}$ ion as well as nanostructurization of VO$_x$
may be crucial for the nucleation of ferromagnetic clusters in
\vox. In this respect one can find a striking similarity with high
temperature ferromagnetism (HTFM) with $T_{\rm c} > 300$\,K
recently observed in nanostructured diluted magnetic
semiconductors (DMS) (see, e.g.,~\cite{Radovanovic,Xing}). Adding
a small percentage of magnetic TM ions to nanocrystals
~\cite{Radovanovic} or nanowires ~\cite{Xing} of nonmagnetic ZnO
yields a robust HTFM that was not achieved by doping the bulk ZnO.
The occurrence of structural inhomogeneities on the nanometer
scale concomitant with the charge localization are the key
prerequisites for this remarkable effect. For example, in Ni:ZnO
nanocrystals, just by tuning the aggregation of nanocrystals, one
obtains HTFM with a $T$-independent saturation value and a small
coercivity, very similar to our findings~\cite{Radovanovic}. The
stabilization of HTFM in DMS has been discussed theoretically in
terms of collective polaronic effects~\cite{Kaminski,Durst},
namely that bound interacting ferromagnetic polarons may be formed
due to exchange interaction of localized charges with magnetic
impurities, in particular in the presence of defects. One can
conjecture a possible relevance of this scenario to \vox\ in view
of some apparent similarities with DMS: (i) - the current-voltage
characteristics of individual tubes reveals a semiconducting
behavior with conductivity decreasing upon Li doping
~\cite{Krusin}; (ii) - electron doping due to the Li intercalation
creates additional spin centers and (iii) - locally distorts the
structure. A delicate balance between these factors controlled by
the Li intercalation may be the reason for a strong sensitivity of
the observed effect to the Li content. At small doping levels the
amount of nonmagnetic $V^{5+}$ ions (which are the 'holes' in the
magnetic subsystem) is big enough to prevent the formation of spin
clusters. On the other hand at large Li dopings \vox\ turn to a
uniform rolled up spin-1/2 plane with predominantly
antiferromagnetic interactions which could be much less sensitive
to a perturbing influence of Li-caused defects.

In summary, we have studied by means of NMR, \musr\ and ESR
spectroscopies combined with static magnetic measurements the
influence of Li intercalation on the magnetic properties of
Li$_{y}$VO$_{x}$-nanotubes. We find a particular concentration of
the Li dopant which turns this compound into a strongly magnetic
material exhibiting ferromagnetism on the room temperature scale.
The data give evidence that this very unusual for a low-spin
vanadium oxide behavior is due to the formation of nanosize
interacting ferromagnetic spin clusters around intercalated Li
ions. Such clusters behave as an ensemble of superparamagnetic
particles with a broad size distribution whose big magnetic
moments can be easily aligned by a moderate magnetic field even at
room temperature. The robustness of the ferromagnetic spin
structure may be suggestive of its collective polaronic nature.

\acknowledgments Support from the DFG (KL 1824/2, 436 RUS 113/936/0-1) and of the RFBR (08-02-91952-NNIO-a, 07-02-01184-a) is gratefully
acknowledged. YCA acknowledges support of the Programme Alban, the European Union Programme of High Level Scholarships for Latin America, scholarship
No. E04D049329CO.


\begin{thebibliography}{99}
\bibitem{fraction}  Due to the electrochemical doping process, $\sim$17\% of the sample mass is paramagnetic carbon which contributes only
to the slowly relaxing signal.
\end{thebibliography}
\end{document}